\documentclass[nofootinbib,showpacs,aps,prc,twocolumn,preprintnumbers,letterpaper]{revtex4-2}
\usepackage{amsmath,amssymb,xspace,braket,float,tabularx}
\usepackage{epsfig}
\usepackage{graphicx}
\usepackage{slashed}
\usepackage{amsfonts}
\usepackage{epstopdf}
\usepackage{tablefootnote}
\usepackage[normalem]{ulem}

\usepackage{subfigure}
\usepackage{hhline}

\usepackage[pdftex, pdfborder= 0 0 0, citecolor=blue, urlcolor=blue, linkcolor=blue, colorlinks=true, bookmarksopen=true]{hyperref}

\newcommand{\gsf}{$\gamma$SF\xspace}
\newcommand{\gsfs}{$\gamma$SFs\xspace}

\newcommand{\ele}[2]{$^{#2}$#1\xspace}
\newcommand{\ssec}[1]{\emph{#1}.---}

\begin{document}

\title{Low-energy enhancement in the magnetic dipole radiation of actinide nuclei}

\author{C. Rodgers, D. DeMartini, and Y. Alhassid}
\affiliation{Center for Theoretical Physics, Sloane Physics Laboratory, Yale University, New Haven, Connecticut 06520, USA}

\begin{abstract}
We present the first theoretical results of the magnetic dipole (M1) $\gamma$-ray strength function (\gsf) for actinide nuclei using the shell-model Monte Carlo (SMMC) method. We observe a low-energy enhancement (LEE) in the M1 \gsfs of the six actinide nuclei studied here, which represents the first evidence, theoretical or experimental, that the LEE persists in the actinides. We also identify a scissors mode resonance in all six nuclei, which we compare with recent Oslo-method experiments. 
\end{abstract}


\maketitle

\ssec{Introduction} \label{sec:intro}
The $\gamma$-ray strength function (\gsf) of a nucleus, which describes the probability of $\gamma$-ray emission and absorption, is a critical input to neutron-capture rate calculations within the Hauser-Feshbach theory~\cite{Hauser1952}. These reaction rates are important for determining elemental abundances from r-process nucleosynthesis~\cite{Mumpower2016}. Furthermore, structures seen in \gsfs  provide insight into the collective behavior of nucleons~\cite{Heyde2010}. Thus, accurate modeling of nuclear \gsfs is important for developing more precise predictions of elemental abundances in stars as well as improving our understanding of nuclear structure. 

The giant dipole resonance, which is an electric dipole (E1) collective mode~\cite{Goldhaber1948}, dominates the \gsf at higher $\gamma$-ray energies.
Nuclei also exhibit resonances at lower energies (below the neutron separation energy $S_n$), some of which are magnetic dipole (M1) in nature. The most prominent M1 modes are the scissors mode resonance (SR)~\cite{Lo1979,Bohle1984,Heyde2010}, seen in deformed nuclei, and the spin-flip mode~\cite{Heyde2010}. Accounting for these low-energy structures in the \gsf of neutron-rich nuclei significantly improves calculations of neutron radiative capture rates~\cite{Mumpower2017}. 

Most experimental measurements of the \gsf rely on either nuclear resonance fluorescence (NRF) reactions~\cite{Berg1987} or the Oslo method~\cite{Schiller2000}, with the latter allowing for simultaneous determination of both the \gsf and the nuclear level density. Experiments on light- and medium-mass nuclei observed a then-unexpected `upbend' in the \gsf at the lowest $\gamma$-ray energies~\cite{Voinov2004,Guttormsen2005,Wiedeking2012,Larsen2013,Larsen2018}, now known as the low-energy enhancement (LEE). Configuration-interaction (CI) shell-model calculations suggested that the LEE is magnetic dipole (M1) in nature~~\cite{Schwengner2013,Brown2014,Schwengner2017,Sieja2017,Karampagia2017,Sieja2018}.

More recently, a LEE has been observed experimentally~\cite{Kheswa2015,Naqvi2019,Simon2016,Guttormsen2022} and calculated theoretically~\cite{Fanto2024,Mercenne2024,DeMartini2025} in the M1 \gsf in chains of neodymium and samarium isotopes. Due to the large shell model spaces associated with these heavy open-shell nuclei,  CI shell-model  calculations are only possible using the shell-model Monte Carlo (SMMC) method~\cite{Johnson1992,Lang1993,Alhassid1994,Koonin1997,Alhassid2001,Alhassid2017}. The SMMC has been used extensively for calculating properties of lanthanide nuclei at finite temperature~\cite{Alhassid2008,Ozen2013, Alhassid2014,Guttormsen2021}. Recently, the SMMC has been extended to calculate the level densities of actinide nuclei~\cite{DeMartini2025b}.

The possible existence of a LEE in actinide nuclei is especially important, as neutron capture rates in these actinides are relevant not only for the astrophysical r-process, but also in processes occurring inside of nuclear reactors~\cite{osti1036581}. If the LEE persists in neutron-rich heavy nuclei, this would have profound effects on $r$-process nucleosynthesis through the enhancement of neutron-capture rates near the neutron drip line~\cite{Larsen2010}.

While various experimental measurements were made of the \gsf in actinide nuclei, many of these experiments utilized NRF reactions~\cite{Adekola2011,Heil1988,Margraf1990,Yevetska2010}.  The LEE has been shown to only exist in the \gsf built on top of excited states. Thus, it cannot be measured in NRF experiments. A LEE can in principle be observed in Oslo-method experiments measuring the decay of excited nuclei. Oslo-method experiments, however, are currently limited to studying actinides down to $\gamma$-ray energies of $E_\gamma \gtrapprox 1\,$MeV~\cite{Guttormsen2012,Guttormsen2014,Laplace2016,Tornyi2014,Garrote2022,Zeiser2019}. While this is sufficient for observing the scissors mode, probing the LEE may require measurements of the \gsf at even lower $\gamma$-ray energies. Thus, the existence of a LEE in the actinides has yet to be experimentally confirmed. 

Here we present the first SMMC calculations of the M1 \gsf at the neutron-separation energy of six actinides that have recently been studied with the Olso method~\cite{Guttormsen2014,Laplace2016,Zeiser2019}: \ele{Th}{232}, \ele{U}{237-239}, \ele{Pu}{240} and \ele{Pu}{243}. We clearly identify a LEE and observe both the scissors and spin-flip modes in all six nuclei. 

We first provide an overview of the computational methods we use to calculate the imaginary-time M1 response function and its conversion into a \gsf using the maximum entropy method~\cite{Fanto2024,Mercenne2024,DeMartini2025}. Then we present results of our calculations in the above six actinide nuclei and compare them with experimental data~\cite{Guttormsen2014,Laplace2016,Zeiser2019}. 

\ssec{Strength and response functions} \label{sec:gsf}
For a nucleus at temperature $T$, we define the strength function of the M1 spherical vector operator $\hat{\mathcal{O}}_{M1}$ as

\begin{align}
    \label{eq:formalSF_spinIndependent}
       S_{M1}(T;\omega) = \sum_{\alpha_i,J_i\alpha_f,J_f} &\frac{e^{-\beta E_{\alpha_i,J_i}}}{Z}|(\alpha_f J_f||\hat{\mathcal{O}}_{M1}||\alpha_i J_i)|^2\\
       &\times \delta(E_{\alpha_f,J_f}-E_{\alpha_i,J_i}-\omega) \nonumber
\end{align}
where $Z = \sum_{\alpha_i,J_i} (2J_i+1)e^{-\beta E_{\alpha_i,J_i}}$ is the canonical partition function and $\hat{\mathcal{O}}_{M1}$  is the magnetic dipole transition operator  defined by
\begin{equation}
    \label{eq:M1op}
    \hat{\mathcal{O}}_{M1} = \sqrt{\frac{3}{4\pi}}\frac{\mu_N}{\hbar c}\left(g_l \hat{\mathbf{l}}+g_s \hat{\mathbf{s}}\right) \;,
\end{equation}
where $\hat{\mathbf{l}}$ and $\hat{\mathbf{s}}$ are, respectively, the orbital and spin angular momentum operators. We use the free-nucleon $g$-factors $g_{l,p} = 1$, $g_{l,n} = 0$, $g_{s,p}  = 5.5857$, and $g_{s,n} = -3.8263$. 

Oslo-method experiments do not measure directly the strength function but rather the de-excitation \gsf. For M1 transitions, we denote the corresponding de-excitation \gsf  by $f_{M1}$; it is defined by~\cite{Bartholomew1973}
\begin{equation}
    \label{eq:fm1}
    f_{M1}(E_i;E_\gamma) = \Tilde{\rho}(E_i)E_\gamma^{-3}\braket{\Gamma_{M1}(E_i;E_\gamma)} \;,
\end{equation}
where $\braket{\Gamma_{M1}(E_i;E_\gamma)}$ is the average partial width for a nucleus at initial energy $E_i$ to emit M1 radiation with energy $E_{\gamma}$, and $\tilde{\rho}$ is the average level density.

The de-excitation strength function $f_{M1}$ is related to the strength function $S_{M1}$ by~\cite{Fanto2024,DeMartini2025})
\begin{align}
    \label{eq:fm1_2}
    f_{M1}(E_i;E_{\gamma}) \approx  a \frac{\tilde{\rho}(E_i)}{3\tilde{\rho}(E_f)} \Bar{S}_{M1}(T;\omega=-E_\gamma) \;,
\end{align}
where $\Bar{S}_{M1}(T)$ is the smoothed analog of Eq.~(\ref{eq:formalSF_spinIndependent}), obtained by integrating over a finite energy resolution $\Delta \omega$, and $a = \frac{16\pi}{9(\hbar c)^3}$.

The \gsf cannot be directly calculated using the SMMC. Instead, we must calculate the imaginary-time response function
\begin{equation} \label{eq:imagTimeResponse}
	R_{M1}(T; \tau) = \langle \hat{\mathcal{O}}_{M1}(\tau) \cdot \hat{\mathcal{O}}_{M1}(0) \rangle\;,
\end{equation}
where $\braket{\ldots}$ denotes a thermal average at temperature $T$, $\hat{\mathcal{O}}_{M1}(\tau) \cdot \hat{\mathcal{O}}_{M1}(0) = \sum_\mu (-)^\mu \hat{\mathcal{O}}^\mu_{M1}(\tau) \hat {\mathcal{O}}^{-\mu}_{M1}(0)$ and $\hat{\mathcal{O}}^\mu_{M1}(\tau) = e^{\tau \hat H}\hat{\mathcal{O}}^\mu_{M1} e^{-\tau \hat H}$  ($\hat{H}$ is the nuclear Hamiltonian).

The strength function in Eq.~(\ref{eq:formalSF_spinIndependent}) is related to the response function with
\begin{equation}
\label{eq:imaginaryTimeWithStrength}
    R_{M1}(T;\tau) = \int_0^\infty d\omega K(\tau,\omega) S_{M1}(T;\omega) \;,
\end{equation}
where $K(\tau,\omega) = e^{-\tau \omega}+e^{-(\beta-\tau)\omega}$ and we have used the symmetry relation $S_{\hat{\mathcal{O}}}(T;-\omega) = e^{-\beta \omega}S_{\hat{\mathcal{O}}}(T;\omega)$. Calculating the $M1$ \gsf thus amounts to inverting Eq.~(\ref{eq:imaginaryTimeWithStrength}).

\ssec{Computational Methods} \label{sec:comp}
We calculate the response function~(\ref{eq:imagTimeResponse}) in the SMMC method~\cite{Alhassid2017}, which is based on the Hubbard-Stratonovich transformation~\cite{Stratonovich1957,Hubbard1959} to express the imaginary-time propagator $e^{-\beta \hat H}$ as a superposition of one-body propagators describing nucleons moving in external axillary fields $\sigma_\alpha$. 
We use the canonical ensemble of fixed numbers of protons and neutrons by applying particle-number projection~\cite{Ormand1994}. The projection on an odd number of protons or neutrons results in a sign problem at low temperatures. However, near the neutron resonance energy the Monte-Carlo sign remains close to 1~\cite{Alhassid2024,DeMartini2025}. 

In the SMMC, the imaginary-time response function~(\ref{eq:imagTimeResponse}) is given by
\begin{equation}
    \label{eq:responseInSMMC}
    R_{M1}(T;\tau) = \frac{\int D[\sigma]W_\sigma \Phi_\sigma \braket{\hat{O}_{M1}(\tau)\cdot\hat{O}_{M1}(0)}_\sigma}{\int D[\sigma]D[\sigma]W_\sigma \Phi_\sigma}
\end{equation}
where $\hat{O}_{M1}(\tau)\cdot\hat{O}_{M1}(0) = \sum_{\mu} (-)^\mu \hat{O}_{M1}^\mu(\tau)\hat{O}_{M1}^{-\mu}(0)$ and $\hat{O}_{M1}(\tau) = \hat{U}_\sigma^{-1}(\tau)\hat{O}_{M1}\hat{U}_\sigma(\tau)$ with $\hat U_\sigma(\tau,0)$ being the propagator between times $0$ and $\tau$ for a given sample $\sigma$ of the auxiliary fields. The expectation value $\langle \ldots \rangle_\sigma$ is taken with respect to $\hat U_\sigma(\beta,0)$, $W_\sigma = G_\sigma |{\rm Tr} \hat U_\sigma(\beta,0)|$ is a positive-definite weight (with $G_\sigma$ a Gaussian factor), and $\Phi_\sigma= {\rm Tr} \hat U_\sigma(\beta,0)/|{\rm Tr} \hat U_\sigma(\beta,0)|$ is the Monte Carlo sign. 
The integration in Eq.~(\ref{eq:responseInSMMC}) is carried out using a Metropolis Monte Carlo algorithm~\cite{Alhassid2017}.

For our calculations of the actinides, we use a valence space comprised of the $0h_{9/2}$, $1f_{7/2}$, $1f_{5/2}$, $2p_{3/2}$, $2p_{1/2}$, $0i_{13/2}$ (82-126 shell)  plus  $1g_{9/2}$ orbitals for protons and $1g_{9/2}, 2d_{5/2}, 0i_{11/2}, 1g_{7/2}, 3s_{1/2}, 2d_{3/2}, 0j_{15/2}$ (126-184 shell)  plus the $1h_{11/2}$ orbitals for neutrons. These were determined by considering orbital occupation numbers, as in previous SMMC analyses~\cite{Alhassid2008}. 
	
The effective residual interactions, which consist of monopole pairing and multipole terms (quadrupole, octupole and hexadecupole), are the same as those used in Ref.~\cite{DeMartini2025b}. 

We compute the imaginary-time response function $R_{M1}(T;\tau)$ via Eq.~(\ref{eq:responseInSMMC}). Inverting Eq.~(\ref{eq:imaginaryTimeWithStrength}) to determine the strength function $S_{M1}(T;\omega)$ is equivalent to numerically inverting a bilateral Laplace transform, which is known to be an ill-posed problem. While a number of methods exist to perform this task~\cite{Tripolt2019}, we use the maximum entropy method (MEM)~\cite{Gubernatis1991,Jarrell1996,Gubernatis_book}, which has been used previously in SMMC calculations of strength functions of heavy nuclei~\cite{Fanto2024,Mercenne2024,DeMartini2025}.

The MEM is a Bayesian approach to maximizing the objective function 
\begin{equation}
\label{eq:objFunc}
    Q(S_{M1},\alpha) = \alpha \mathcal{S}-\frac{1}{2}\chi^2
\end{equation}
with respect to $S_{M1}$. Here, $\chi^2$ is the likelihood function given by 
\begin{equation}
    \chi^2 = (R^{\mathrm{SMMC}}_{M1}-R_{M1})^TC^{-1}(R^{\mathrm{SMMC}}_{M1}-R_{M1})
\end{equation}
where $C$ is the covariance matrix of the SMMC response function data $R^{\mathrm{SMMC}}_{M1}$, and $R_{M1}$ is calculated using Eq.~(\ref{eq:imaginaryTimeWithStrength}) for a given choice of $S_{M1}$. The entropy function $\mathcal S$ in Eq.~(\ref{eq:objFunc}) is defined by
\begin{equation}
    \mathcal{S} = \int d\omega \left(S_{M1}-S^{\mathrm{prior}}_{M1}-S_{M1}\log{\left(S_{M1}/S_{M1}^{\mathrm{prior}}\right)} \right) \;,
\end{equation}
where $S^{\mathrm{prior}}$ is a prior distribution of the strength function. 

\begin{figure*}[bth]
	\includegraphics[width=\linewidth]{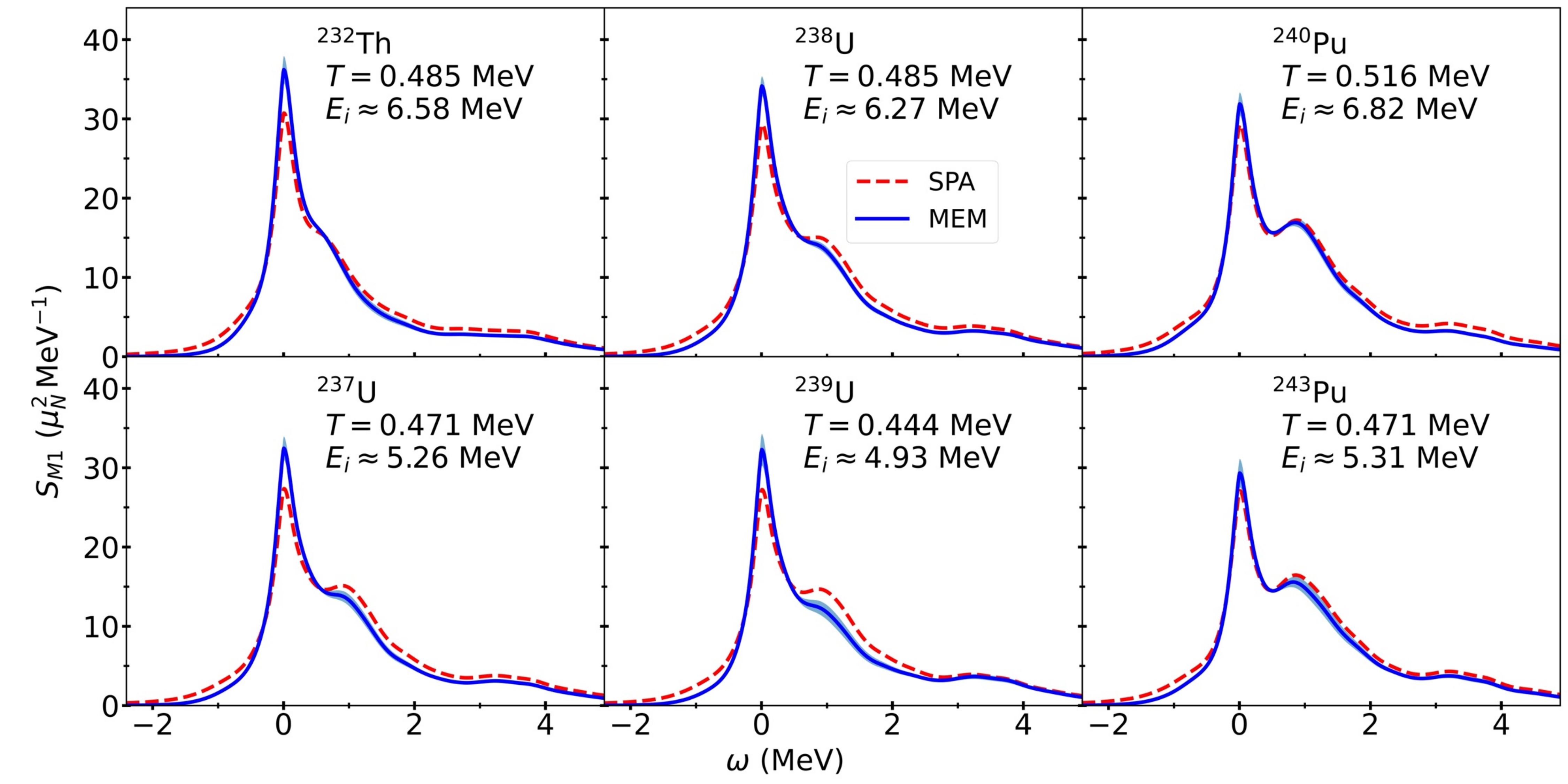}
	\caption{The M1 strength functions $S_{M1}$ for six actinides at temperatures near their neutron separation energy. The blue solid lines are the SMMC+MEM results with shaded blue bands representing the uncertainties, and the red dashed lines are the SPA M1 strength functions used as the prior.}
	\label{fig:nsep_sf}
\end{figure*}

To account for the parameter $\alpha$, we use Bryan's method~\cite{Bryan1990} which averages the solutions $S_{M1}$ that maximize Eq.~(\ref{eq:objFunc}) over all values of $\alpha$.

The success of the MEM is depends on a suitable choice of the prior distribution. For this analysis, we will take our prior to be the static-phase approximation (SPA)~\cite{Puddu1991,Lauritzen1988,Alhassid1992,Rossignoli1998,Rossignoli1999}, as previous theoretical studies of the lanthanides demonstrated that the SPA provides a good approximation for the M1 response function at temperatures near the neutron separation energy~\cite{Fanto2024,Mercenne2024,DeMartini2025}. This choice for the prior strength function is confirmed for actinides in the Supplemental Material by comparing its imaginary-time response function with the SMMC response function~\cite{suppmat}.

\ssec{Results for actinides} \label{sec:res}
We calculated the $M1$ strength functions $S_{M1}$ of six actinide nuclei: three even-even   \ele{Th}{232}, \ele{U}{238}, and \ele{Pu}{240} and three odd-even   \ele{U}{237},  \ele{U}{239}, and \ele{Pu}{243} near their neutron separation energy. The results are shown in Fig.~\ref{fig:nsep_sf}. For both the SMMC and SPA calculations we used 8,000 uncorrelated samples, and uncertainties were calculated using jackknife resampling.

We clearly see a large peak  at $\omega = 0$ in all six nuclei, which we identify as the LEE (see below).  We also identify two more peaks at $\omega \approx 1$\,MeV and $\omega \approx 3.5$\,MeV. We interpret these, respectively, as the scissors mode and spin-flip mode built on top of excited states. In the Supplemental Material, we also calculate the $M1$ $\gamma$SF of the six actinides at low temperatures close to the ground state~\cite{suppmat} and observe that the LEE peak is missing. Both the scissors mode and spin-flip modes are observed in the $M1$ strength function computed near the ground state and show more structure when compared with the $M1$ strength function computed near the neutron resonance energy.

\begin{figure*}[bth]
    \includegraphics[width=\linewidth]{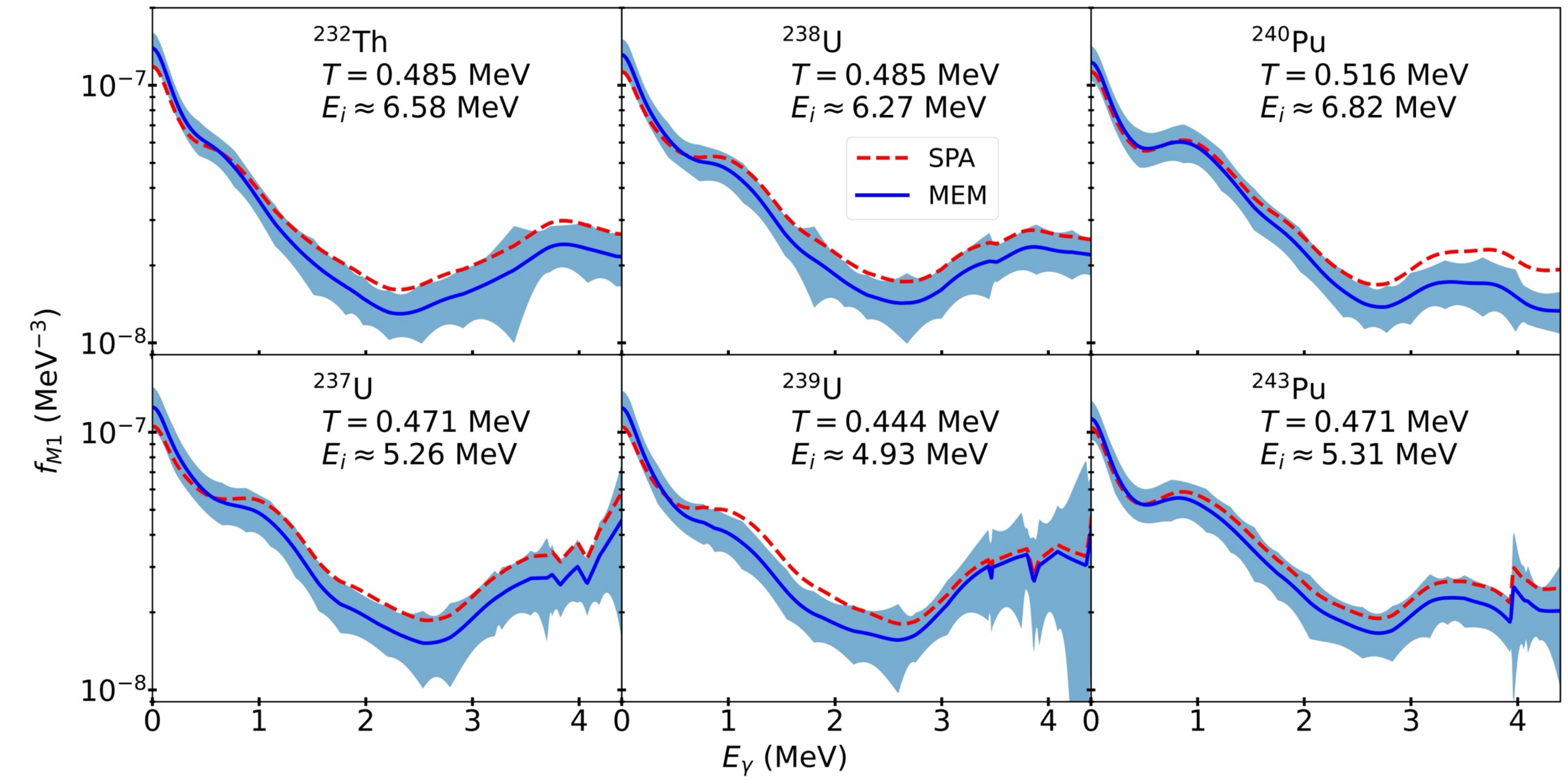}
    \caption{The de-excitation strength functions $f_{M1}$ calculated from Eq.~(\ref{eq:fm1_2}) for the six actinides in fig.~\ref{fig:nsep_sf}, calculated near their neutron separation energies. The blue solid lines are the SMMC+MEM results with shaded blue bands representing the uncertainties, and the red dashed lines describe the SPA priors.}
    \label{fig:gamma_sf}
\end{figure*}

We convert $S_{M1}$ into de-excitation strength functions $f_{M1}$ using Eq.~(\ref{eq:fm1_2}), and the results are shown in Fig.~\ref{fig:gamma_sf}. The level densities $\tilde{\rho}$ in Eq.~(\ref{eq:fm1_2}) were calculated with the SMMC as in Ref.~\cite{DeMartini2025b}. 

As suggested by Refs.~\cite{Schwengner2013,Karampagia2017} we fit the LEE to an exponential form

\begin{equation}\label{eq:LEE_fit}
    f^{LEE}_{M1}(E_\gamma) = C_{0}e^{-\kappa E_\gamma}.
\end{equation}
The LEE fit parameters $C_0$ and $\kappa$ for all six actinide nuclei care listed in Table~\ref{tab:LEE_fit}.

\begin{figure}
    \subfigure{\includegraphics[width=\linewidth]{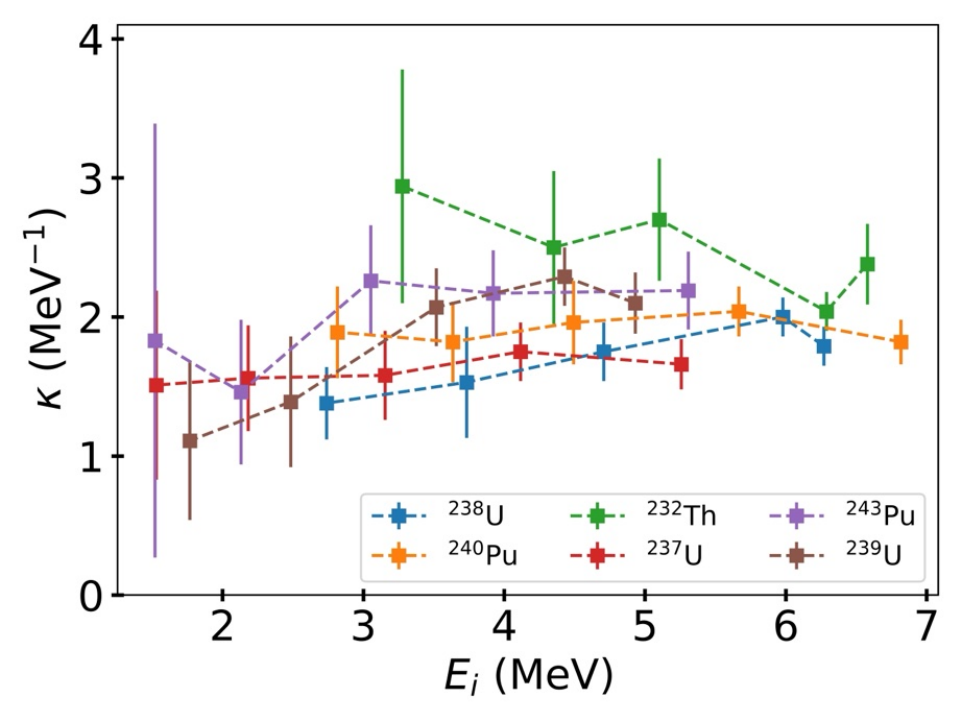}}
   
    \caption{The slope parameter $\kappa$, obtained by fitting the LEE to Eq.~\eqref{eq:LEE_fit}, as a function of the initial energy $E_i$ for all six actinide nuclei.}
    \label{fig:lee_temp_dep}
  \end{figure}

\begin{table*} [htb]
\centering
	\caption{Parameters for the fits of the $M1$ \gsf to the form in Eq. (\ref{eq:LEE_fit})} 
\begin{ruledtabular}
	\begin{tabular}{c|cccccc}
Nucleus & \ele{Th}{232} & \,\, \ele{U}{237} & \,\, \ele{U}{238} & \,\, \ele{U}{239} & \,\, \ele{Pu}{240} & \,\, \ele{Pu}{243} \\ \hline \\[-0.3cm]
		$C_0 \,\, (10^{-7}$\,MeV$^{-3}$) &\,\, 1.48 (1) &\,\,\,\, 1.38 (6) &\,\,\,\, 1.32 (6)  &\,\,\,\, 1.28 (7) &\,\,\, 1.26 (1) &\,\,\,\, 1.19 (7)\\[0.1cm] \hline \\[-0.3cm]
        $\kappa$ (MeV$^{-1}$) &\,\, 2.38 (29) &\,\,\,\, 1.66 (18) &\,\,\,\, 1.79 (14) &\,\,\,\, 2.10 (22) &\,\,\,\, 1.82 (16) &\,\,\,\, 2.19 (28)
	\end{tabular}
        \end{ruledtabular}
	\label{tab:LEE_fit}
\end{table*}
In Fig.~\ref{fig:lee_temp_dep} we show the fitted values of $\kappa$ as a function of the initial energy $E_i$. We find that $\kappa$ depends weakly on $E_i$. This is consistent with previous results from SMMC~\cite{DeMartini2025,Fanto2024,Mercenne2024} and CI shell-model~\cite{Karampagia2017} calculations in lanthanide and mid-mass nuclei, respectively.

Subtracting the LEE fit~(\ref{eq:LEE_fit}) from our $f_{M1}$ results allows us to isolate the other modes in the M1 strength. In particular, we fit the SR in the residual strength to a Lorentzian form~\cite{Guttormsen2014}
\begin{equation}\label{eq:sr_lorentz}
	f_{M1}^{\textrm{SR}}(E_{\gamma}) = \frac{1}{3 \pi^2 (\hbar c)^2} \frac{\sigma E_{\gamma} \Gamma^2}{(E_{\gamma}^2 - \omega^2)^2 + E_{\gamma}^2 \Gamma^2} \;,
\end{equation}
where $\sigma$, $\omega$, and $\Gamma$ are, respectively, the strength, center, and width of the SR. 

We note that the SRs measured in Ref.~\cite{Guttormsen2014,Laplace2016} are actually comprised of two lorentzians, whereas in our calculations we see only a single peak (a possible reason for this discrepancy will be discussed below). Thus, rather than comparing the fit parameters, we compare only the total integrated strength

\begin{equation}
\label{eq:bm1}
B(M1) = \frac{27(\hbar c)^3}{16 \pi} \int_0^{S_n} f_{M1}(E_{\gamma}) dE_{\gamma} \;.
\end{equation}

Our calculated integrated strengths and the experimentally measured values from~\cite{Guttormsen2014,Laplace2016} are tabulated in Table~\ref{tab:sr_fit}. For completeness we also include the parameters of our Lorentzian fits~(\ref{eq:sr_lorentz}).

\begin{table}
\centering
	\caption{Fit parameters of the M1 SR fitted to the Lorentzian form of Eq.~(\ref{eq:sr_lorentz}) and their integrated strengths (\ref{eq:bm1}) compared to experiments~\cite{Guttormsen2014,Laplace2016}.} 
        \begin{ruledtabular}          
        \begin{tabular}{c|cccc|c}
		Nucleus & \multicolumn{4}{c|}{SMMC + MEM} & \multicolumn{1}{c}{Experiment} \\
		& $\omega$ & $\sigma$ & $\Gamma$ & $B(M1)$  &  $B(M1)$   \\
		& (MeV) & (mb) & (MeV) & $(\mu_N^2)$ & $(\mu_N^2)$ \\
		\hline \\[-0.3cm]
          \ele{Th}{232} & 1.02 (7) & 0.258 (11) & 1.20 (26) & 10.7 (25) & 9.5 (26) \\
          \ele{U}{237} & 1.26 (4) & 0.346 (18) & 1.14 (17) & 11.1 (18) & 8.8 (15) \\
          \ele{U}{238} & 1.26 (4) & 0.352 (14) & 1.14 (18) & 11.3 (19) & 9.4 (16)\\
          \ele{U}{239} & 1.32 (5) & 0.362 (14) & 1.37 (21) & 13.2 (25) & 8.8 (14) \\
          \ele{Pu}{240} & 1.30 (3) & 0.504 (16) & 1.40 (14) & 19.2 (21) & -\\
          \ele{Pu}{243} & 1.31 (5) & 0.557 (17) & 1.48 (19) & 22.2 (30) & 10.1 (15) 
        \end{tabular}
        \end{ruledtabular}          
        
	\label{tab:sr_fit}
\end{table}

To assess the accuracy of our theoretical values for $B(M1)$ when compared to the experimental measurements, we note several key points. First, the experimental measurements do not distinguish the parity of the $\gamma$-rays, namely the measured strengths contain contributions from both E1 and M1 radiation. While the experimental results account for the presence of the E1 giant dipole resonance by subtracting its extrapolated tail, theoretical calculations have suggested that thermal excitations in the continuum may contribute to an enhancement of low-energy E1 transitions~\cite{Litvinova2013}. Thus there might be residual E1 radiation in the experimental SR results. Second, our calculations are performed at a fixed temperature (canonical ensemble) whereas the experiment is performed at a fixed initial energy $E_i$ (microcanonical ensemble). Finally, the integrated strength in our calculations is sensitive to the $\gamma$-ray energy window used in our lorentzian fit, leading to additional systematic uncertainties in the fit parameters.

Accounting for these factors, we conclude that the integrated strengths of our scissors modes are in reasonable agreement with those measured by Refs.~\cite{Guttormsen2014,Laplace2016}.

The experimental measurements of the SR for the actinides considered here contain two peaks, rather than the single peak we observe in our calculations. Further, the experimental SR peaks are located at a higher energy than our theoretical calculations suggest, as was observed in previous SMMC calculations of the lanthanides~\cite{DeMartini2025}.

It has been suggested that triaxiality can lead to a splitting of M1 modes by an amount proportional to   $\sin{\gamma}$~\cite{Lipparini1989}. Furthermore, because the SR is the counter-rotation of the deformed proton and neutron densities, the energy of the SR mode is positively correlated to the $\beta$ and $\gamma$ deformation of the respective nucleus.

In previous work studying nuclear deformation in the lanthanides~\cite{Mustonen2018} it was found that in the restricted model space used a phenomenological factor accounting for core polarization must be included to generate sufficient deformation in the SMMC. Thus, the SMMC calculations presented here underpredict the deformation of the six actinide nuclei, which could potentially result in the SR having a lower energy than expected and the two SR peaks being too closely split to be resolved.

On the other hand, since the LEE depends only weakly on deformation~\cite{Mercenne2024}, we expect our results for the LEE to be robust. 

\ssec{Summary and discussion} \label{sec:con}
We have performed SMMC calculations of the M1 \gsf for the actinide nuclei \ele{Th}{232}, \ele{U}{237-239}, \ele{Pu}{240}, and \ele{Pu}{243} near their neutron separation energy. To carry out the necessary analytic continuation we used the MEM, with a prior strength function calculated in the SPA.

We observed a  LEE in all six actinides.  These results are the first theoretical prediction of a LEE in the M1 \gsf of actinides. We also observe a scissors mode in all six actinides. 

Improvements to the results presented here could be made by studying the spin dependence of the \gsf. This would allow us to more directly compare our theoretically calculated $f_{M1}$ to experimental measurements, which populate spins differently from the intrinsic spin distribution of the nucleus. 

\ssec{Acknowledgments}
We thank P. Fanto for the use of the SPA and MEM codes he developed. This work was supported in part by the U.S. DOE grant No.~DE-SC0019521. The calculations used resources of the National Energy Research Scientific Computing Center (NERSC), a U.S. Department of Energy Office of Science User Facility operated under Contract No. DE-AC0205CH11231.

The data that support the findings of this article are openly available~\cite{data}.

\bibliography{LEE_actinides.5}
\end{document}


\title{Supplemental Material: \\ Low-energy enhancement in the magnetic dipole radiation of actinide nuclei}

\author{C. Rodgers, D. DeMartini, and Y. Alhassid}
	
\affiliation{Center for Theoretical Physics, Sloane Physics Laboratory, Yale University, New Haven, Connecticut 06520, USA}
	
\maketitle

\subsection{$M1$ $\gamma$SF at low temperatures (near the ground state)}

In this section we compare the magnetic dipole strength function at the neutron resonance energy (discussed in the main text) with the corresponding strength function at low temperatures near the ground state.  

\begin{figure*}[bth]
\centering
\includegraphics[width=0.8\linewidth]{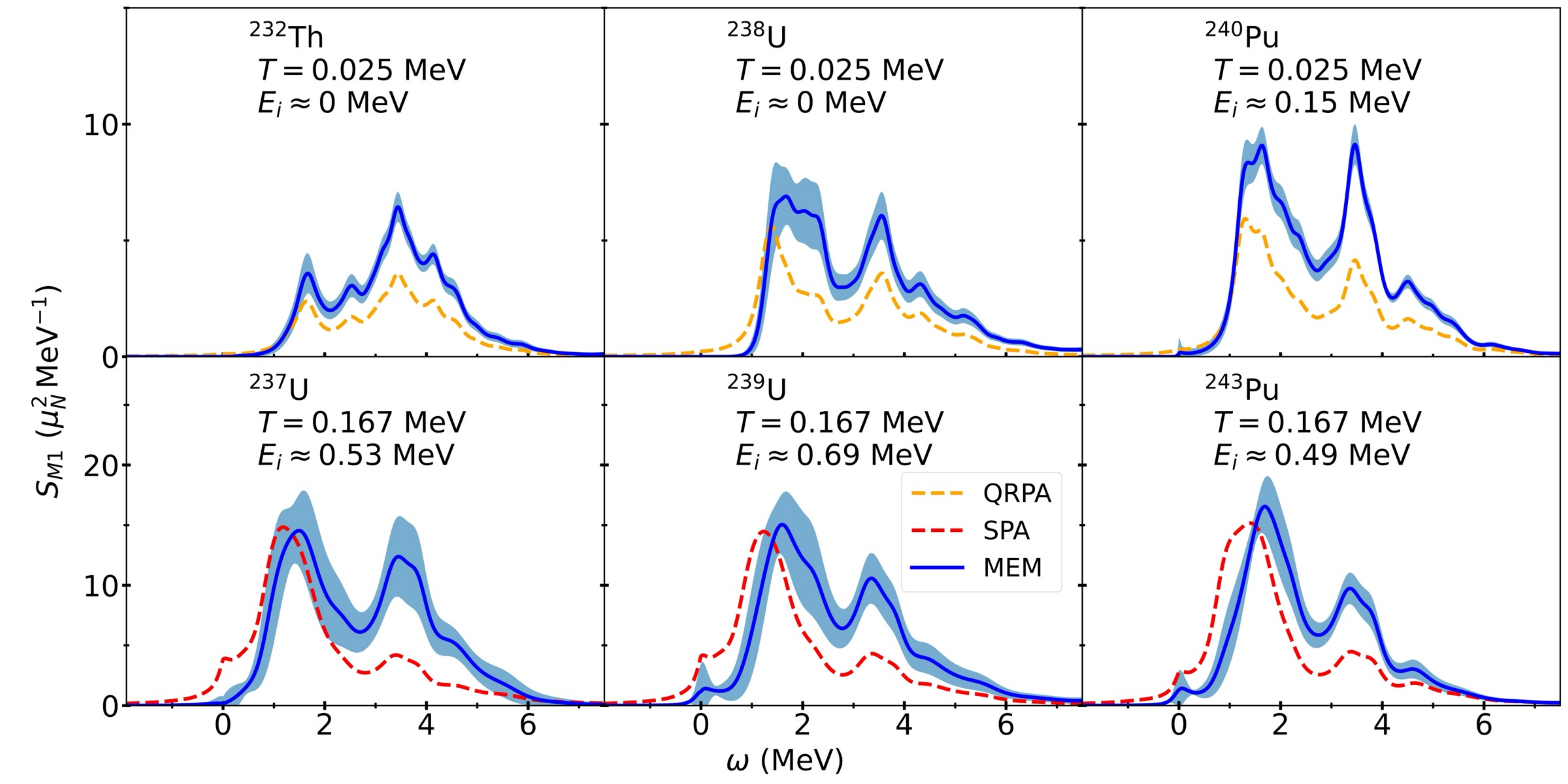}\\
	\caption{The $M1$ \gsf for temperatures near the ground state for the even-mass (top row) and odd-mass (bottom row) actinides. The solid blue lines correspond to the MEM fit result and the blue bands describe the associated statistical uncertainties. The dashed orange lines are the QRPA prior strength functions for the even-mass actinides and the dashed red lines are the SPA prior strength functions for the odd-mass actinides.} 
	\label{fig:gs_sf}
\end{figure*}

 In Fig.~\ref{fig:gs_sf} we show the $M1$ $\gamma$SF $S_{M1}$ for the six actinide nuclei studied in this work: \ele{Th}{232}, \ele{U}{237}, \ele{U}{238}, \ele{U}{239}, \ele{Pu}{240}, and \ele{Pu}{243}. For the even-mass nuclei (top row), we use $T=0.025$ MeV. However, the odd-mass nuclei have a Monte Carlo sign problem at low temperatures  and for these nuclei we use $T=0.167$ MeV (bottom row) for which the sign problem is still moderate~\cite{DeMartini2025}. 

 As discussed in the main text, the \gsfs are calculated using the maximum entropy method (MEM). For very low temperatures we find that the quasiparticle random phase approximation (QRPA) provides a better approximation (see Sec.~\ref{sec:response-functions} below) than the static phase approximation (SPA). Thus for the even-mass nuclei we choose for the prior strength to be the QRPA $M1$ strength function (shown by the dashed orange lines in Fig.~\ref{fig:gs_sf}). For the odd-mass nuclei where the temperature is not so low, we use as prior  the SPA strength function (dashed red lines in Fig.~\ref{fig:gs_sf}). The uncertainties were calculated using jackknife resampling. 

Comparing with the $M1$ $\gamma$SF at the neutron resonance energy, we find that the LEE is absent for the $M1$ $\gamma$SF near the ground state. On the other hand, the scissors mode at $\sim 1.5$ MeV and the spin-flip model at $\sim 3.5$ MeV are more pronounced and show more structure.

\subsection{Imaginary-time response functions}
\label{sec:response-functions}

Using the SMMC, we calculate the exact (up to statistical error) imaginary-time $M1$ response function. We then carry out the numerical analytic continuation in the MEM using a suitable prior strength function. The success of the method depends on a good choice for this prior. In particular, we consider the SPA and QRPA $M1$ strength functions as possible priors.

\begin{figure*}[bth]
  \centering
   \includegraphics[width=1\linewidth]{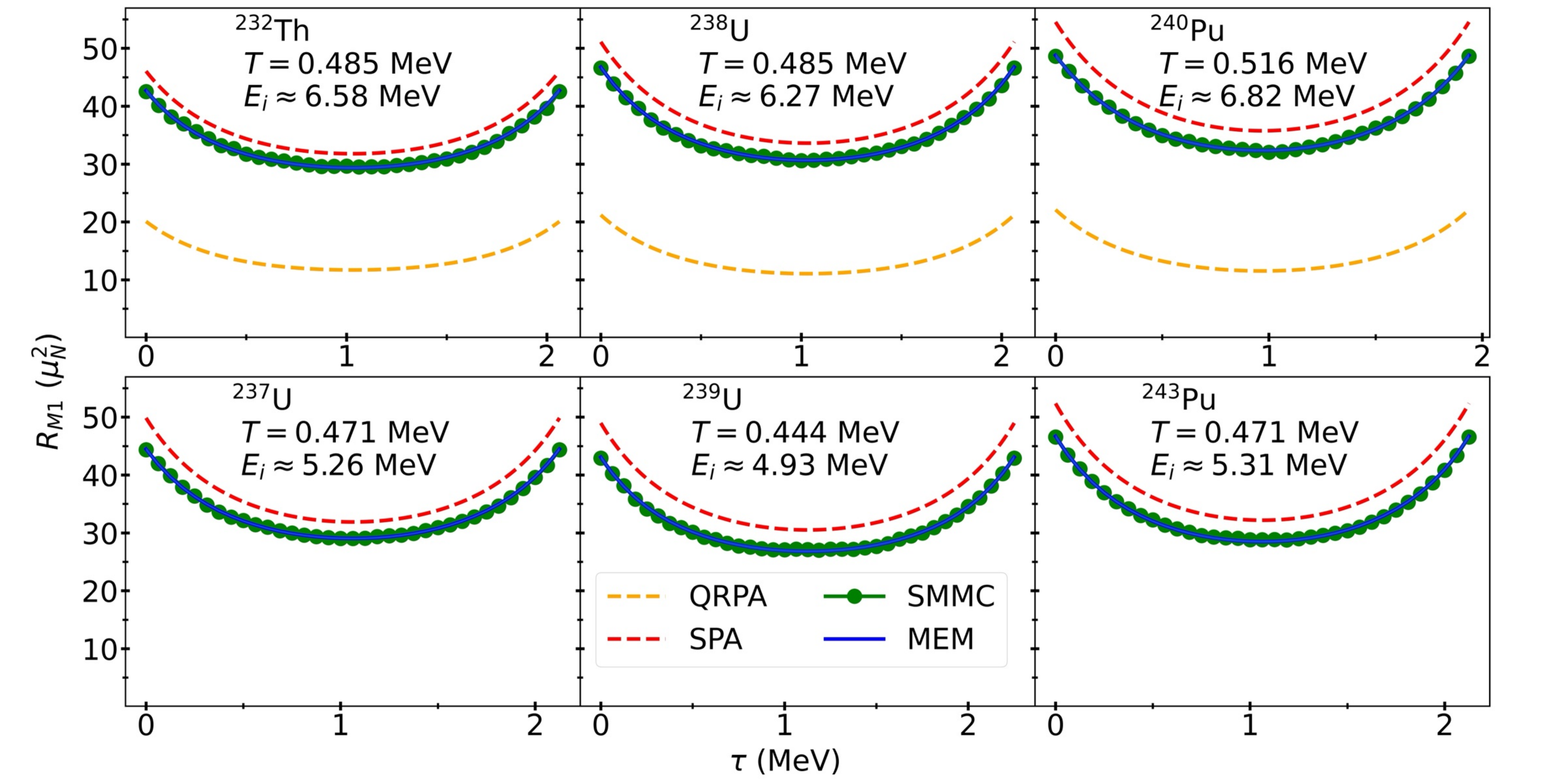}
  	\caption{The finite temperature imaginary-time $M1$ response function for all six actinide nuclei at temperatures near the neutron separation energy. The green circles are the SMMC results, and the solid blue lines are the response functions resulting from the MEM fit. The dashed red lines are the SPA response functions and the dashed orange lines are the QRPA response functions.  For all nuclei we use the SPA strength as the prior for the MEM fit.}
	\label{fig:nsep_rf}
      \end{figure*}

To assess the suitability of different priors, we compare their imaginary-time response functions to the SMMC response. In Fig.~\ref{fig:nsep_rf} we compare the SMMC imaginary-time $M1$ response functions (green circles) for the six actinides at energies near the neutron separation energy to the SPA response (dashed red lines) and the QRPA response (dashed orange lines)  functions. The latter is calculated for the even-mass nuclei only using the code HF-SHELL\footnote{HF-SHELL only supports calculations in even-even nuclei.}~\cite{Ryssens2021}.  We find out that the SPA response function is much closer to the SMMC response function than the QRPA. We therefore use the SPA $M1$ strength function as the prior for temperatures that correspond to the neutron separation energies (for both the even- and odd-mass nuclei). 

As is seen in Fig.~1 of the main text, the LEE is already observed in the SPA $M1$ strength function. This suggests that the large amplitude fluctuations around the mean-field configuration, which are included in the SPA, are important for generating the LEE. 

\begin{figure*}[bth]
	\centering
        \includegraphics[width=1\linewidth]{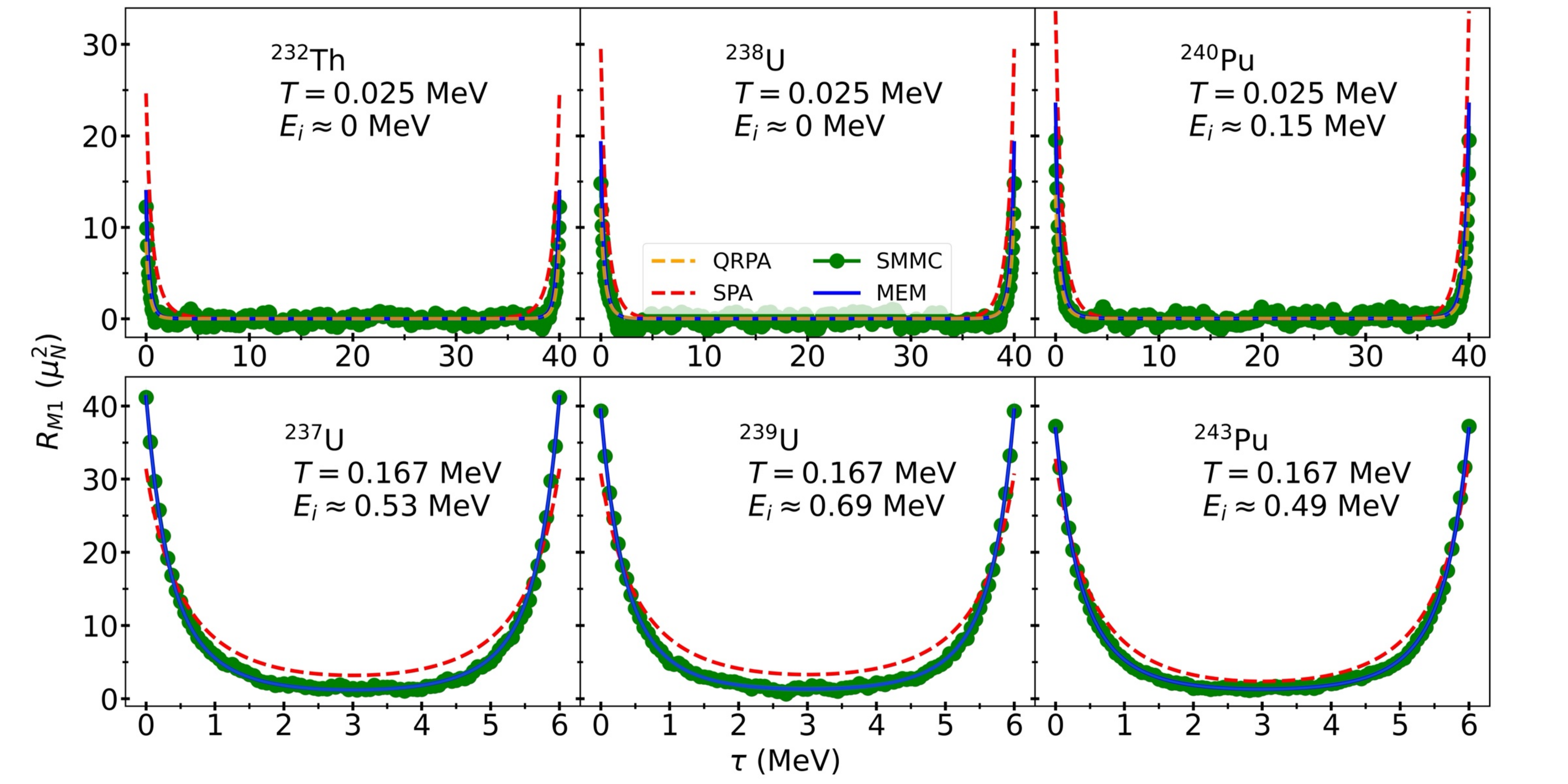}
	\caption{The imaginary-time $M1$ response function for the six actinides at temperatures close to the ground state ($T=0.025$ MeV for the even-mass-nuclei and $T=0.167$ MeV for the odd-mass nuclei). The symbols and various lines are as in  Fig.~\ref{fig:nsep_rf}, except that the QRPA strength is used as a prior for the even-mass nuclei in the top row (while the SPA strength is used as a prior for the odd-mass nuclei in the bottom row).}
	\label{fig:gs_rf}
\end{figure*}

In Fig.~\ref{fig:gs_rf}  we show the $M1$ imaginary-time response functions at low temperatures: $T=0.025$ MeV for the even-mass actinides and $T=0,167$ MeV for the odd-mass actinides. The SMMC results (green circles) are compared with the SPA (dashed red lines) and the QRPA (dashed orange lines) response functions. Based on the results for the even-mass actinides, we conclude that at very low temperatures (close to the ground state), the QRPA results, which account for small amplitude quantal fluctuations about the mean-field configuration are closer to the SMMC results. Therefore, in the MEM for very low temperatures, we use the QRPA strength as the prior strength function. 

\bibliography{LEE_actinides.5}